\documentclass[twocolumn,preprintnumbers,superscriptaddress,amsmath,amssymb,prb]{revtex4}

\usepackage{graphicx}
\usepackage{dcolumn}
\usepackage{subfigure}
\usepackage{bm}

\begin{document}

\title{A first principles study of the stability and mobility of defects in titanium carbide}

\author{Mikael R{\aa}sander}
\email{mikra@kth.se}
\affiliation{%
Department of Physics and Astronomy, Division for Materials Theory,
Uppsala University, PO Box 516, 751 20 Uppsala, Sweden
}%
\affiliation{%
Department of Materials and Nano Physics, KTH Royal Institute of Technology, Electrum 229, 164 40 Kista, Sweden
}%
\author{Biplab Sanyal} 
\affiliation{%
Department of Physics and Astronomy, Division for Materials Theory,
Uppsala University, PO Box 516, 751 20 Uppsala, Sweden
}%
\author{Ulf Jansson}
\affiliation{%
Department of Materials Chemistry,
Uppsala University, PO Box 538, 751 21 Uppsala, Sweden
}%
\author{Olle Eriksson}
\affiliation{%
Department of Physics and Astronomy, Division for Materials Theory,
Uppsala University, PO Box 516, 751 20 Uppsala, Sweden
}%

\date{\today}

\begin{abstract}
We have performed density functional calculations of the formation energies of substitutional transition metal (TM) defects, C vacancies, and C interstitial defects in TiC. In addition we have evaluated the migration energy barriers for C atoms in the presence of TM impurities. We find that the solubility of TM impurities taken from the 3d TM series is low and only Sc and V impurities can be dissolved into TiC at equilibrium. In addition, we find that the migration energy barriers of C in TiC are greatly affected by the presence of TM impurities: The migration barriers are generally lower in the presence of impurities compared to pure TiC and show a clear dependence on the atomic size of the TM impurities. We propose that the mobility of C in TiC will be the highest in the presence of TM impurities from the middle of the 3d TM series. 
\end{abstract}

\maketitle

\section{Introduction}\label{introduction}
Transition metal carbides (TMC) have for a long time been the focus of extensive research. This interest has been spawned by their many interesting physical properties such as high hardness, high melting temperature and high thermal and electric conductivity.\cite{toth} Properties that have given these materials many industrial applications, for example as hard coating materials. Recently, Wilhelmsson et al.\cite{wilhelmsson}  have shown that by selective alloying of TiC with weak carbide forming metals, it is possible to grow solutions of ternary TMC that have a built-in driving force for the release of C from the carbide system. Furthermore, it was proposed that these types of systems could be used as low-friction coating materials.\cite{wilhelmsson} The alloying reduces the otherwise strong covalent metal-to-carbon bonds that are prevailing in TMC, and it is more favorable for C atoms to diffuse to the surface by the creation of C vacancies in the TMC.\cite{rasander,Jansson,Jansson2} In experiments on thin films of nanocrystalline (Ti,M)C, where the carbide phase is embedded in an amorphous C matrix, i.e. (nc-(Ti,M)C/a-C), it has been shown that the alloying with weak carbide forming metals, M, yields a larger amount of C matrix in relation to the carbide phase for systems with identical carbon-to-metal ratio during growth compared to systems that only contain Ti and C.\cite{Jansson,Jansson2,wilhelmsson,rasander,refB4,refG1,refG2} Furthermore, it has been shown that during heat treatment at elevated temperatures there is also an increase of the C matrix phase for the alloyed samples which is not present in the binary systems.\cite{Jansson,Jansson2,rasander} When used as a surface coating material, the ternary solutions yield a lower friction coefficient than the binary samples while maintaining a relatively high degree of hardness.\cite{Jansson,wilhelmsson} In a previous paper we have shown using density functional calculations that release of C is a favorable process for solutions of ternary (Ti,M)C, where the metal M is any metal from the 3d transition metal (TM) series, with the sole exception being V where C release never occurs.\cite{rasander} Furthermore, it was found that it is possible to tune the driving force for the release of C by careful selection of the alloying metal as well as its amount.  However, the details concerning diffusion of C in the ternary solution was not addressed.
\par
Rock salt TiC is a common and often studied TMC.\cite{Haglund1,Haglund2,Isaev1,Isaev2,Vojvodic} Among the 3d TM elements, Ti is involved in the formation of the chemically most stable TMC, since the electronic structure of TiC is such that the Fermi level is positioned in a region with relatively low density of states, which separates bonding from anti-bonding states. TiC is therefore maximally bonding since the occupation of bonding states are complete. By increasing the number of electrons in the system, anti-bonding states will begin to be filled and the stability of the TMC is lowered.\cite{rasander,Haglund1,Haglund2} In fact, there are no stable TMC formed by metals found late in the 3d TM series. Sc, Ti and V are the only 3d TM elements that form TMC in the rock salt structure, while Cr and Mn only form complicated metal-rich TMC similar to Cr$_{23}$C$_{6}$.\cite{toth} Fe, Co and Ni only form meta-stable TMC in the cementite structure, i.e. Fe$_{3}$C.\cite{Meschel} A striking feature of TiC is its defect structure since ideal stoichiometry is almost never found and the variation in carbon-to-metal ratio can be rather large.\cite{toth,Tan,Hugosson2,pasha,Redinger} The mobility of the atoms in TiC has not been extensively investigated by theory and especially so for high C vacancy contents. Tsetseris et al.\cite{Tsetseris} and Razumovskiy et al.\cite{Volya} have investigated the migration energy barriers for a number of diffusion processes in binary TiC in good agreement with experimentally obtained migration energy barriers.\cite{Ettmayer,vanLoo,Kohlstedt,Hoondert}  It is well established, however, that C is the most mobile element in~TiC.
\par
\begin{figure}[t]
\includegraphics[width=5cm]{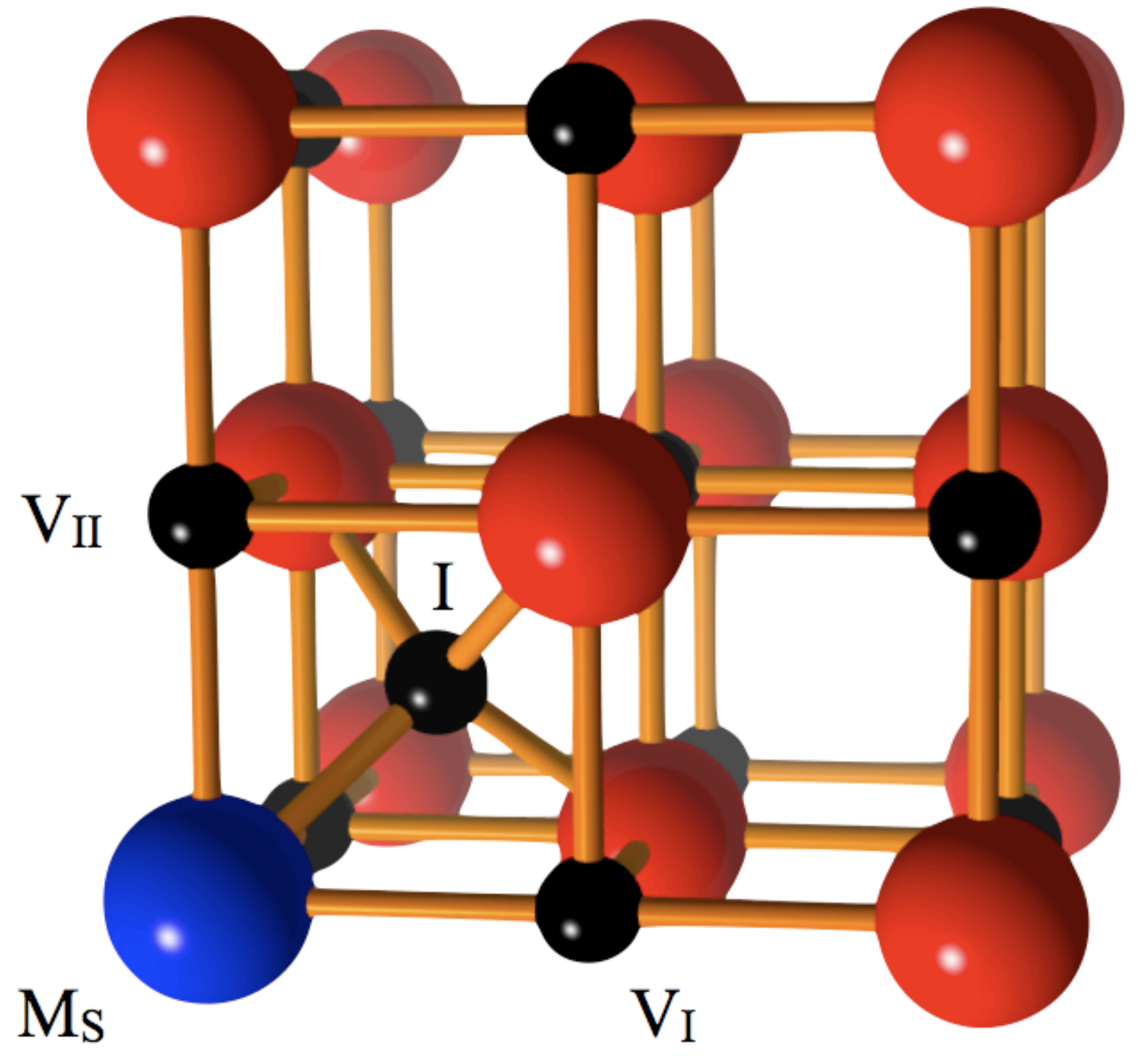}
\caption{\label{fig:defects_ill} (Color online) Crystal structure of rock salt TiC. Ti atoms are symbolized by large (red) spheres and the C atoms by small (black) spheres. A single 3d TM defect is illustrated by the (blue) sphere in the bottom left corner, $M_{S}$. Two C positions have been labelled by $V_{I}$ and $V_{II}$ as is explained in the text. In addition, an interstitial C atom is represented in the tetrahedrally coordinated interstitial position $I$.}
\end{figure}
In this paper, we will investigate the stability of TM impurities in TiC as well as how the stability of C is affected by the presence of these impurities. In addition, we will address the issue of diffusion of C in TiC in the presence of TM impurities. We will show that both the stability of C as well as the migration barriers for C diffusion are greatly affected by the presence of impurities in TiC.
\par
The paper is arranged as follows: In Section \ref{method} we will present the details of the theoretical calculations and in Section \ref{results} our results will be presented. Finally, in Section \ref{conclusions} we will summarize our results and draw conclusions.
\section{Theoretical details}\label{method}
TiC crystallizes in the B1(rock salt) crystal structure which is built up by two intersecting face centered cubic (fcc) lattices where the Ti and C atoms each occupy one of the sublattices. In this study we are mainly interested in four kinds of defects which are illustrated in Fig.~\ref{fig:defects_ill}; namely (a) single 3d TM defects that substitute for Ti, illustrated by M$_{s}$ in Fig.~\ref{fig:defects_ill}, (b) C vacancies created by removing either of the C atoms labelled $V_{I}$ and $V_{II}$ in Fig.~\ref{fig:defects_ill}, and (c) C interstitial defects illustrated by the C atom labelled $I$ in Fig.~\ref{fig:defects_ill} that occupy the tetrahedral position of the metallic lattice. The interstitial defects will be set up in two different ways; the first is to insert an additional C atom into the tetrahedral site of TiC (type I), the second is to move one of the C atoms on the C lattice into the tetrahedral interstitial site leaving a C vacancy on the lattice (type II). This latter type of interstitial defect is more commonly known as a Frenkel defect. C vacancies and C interstitial defects have been evaluated for pure TiC as well as in the presence of TM impurity atoms as is illustrated in Fig.~\ref{fig:defects_ill}.
\par
The different structures illustrated in Fig.~\ref{fig:defects_ill} have been studied using density functional calculations, where we have used the generalized gradient approximation (PW91) for the exchange and correlation energy functional.\cite{perdewandwang91} The Kohn-Sham equation has been solved using Bl{\"o}chls' projector augmented wave method\cite{bloechl} as it is implemented in the Vienna {\it ab-initio} simulation package (VASP).\cite{KresseandFurth,KresseandJoubert} The total energy has been converged with respect to the number of k-points, which has been set up using the special k-points method of Monkhorst and Pack\cite{MonkhorstandPack} and we have used a plane wave energy cut-off of 600 eV. Relaxations of the atomic coordinates have been performed until the Hellmann-Feynman forces acting between different atoms are smaller than 5~meV/{\AA}. Spin-polarized calculations have been used when substitutional defects of Cr, Mn, Fe, Co and Ni have been present. However, the magnetic properties will not be discussed.
\par
\begin{figure}[t]
\centering
\includegraphics[width=3cm]{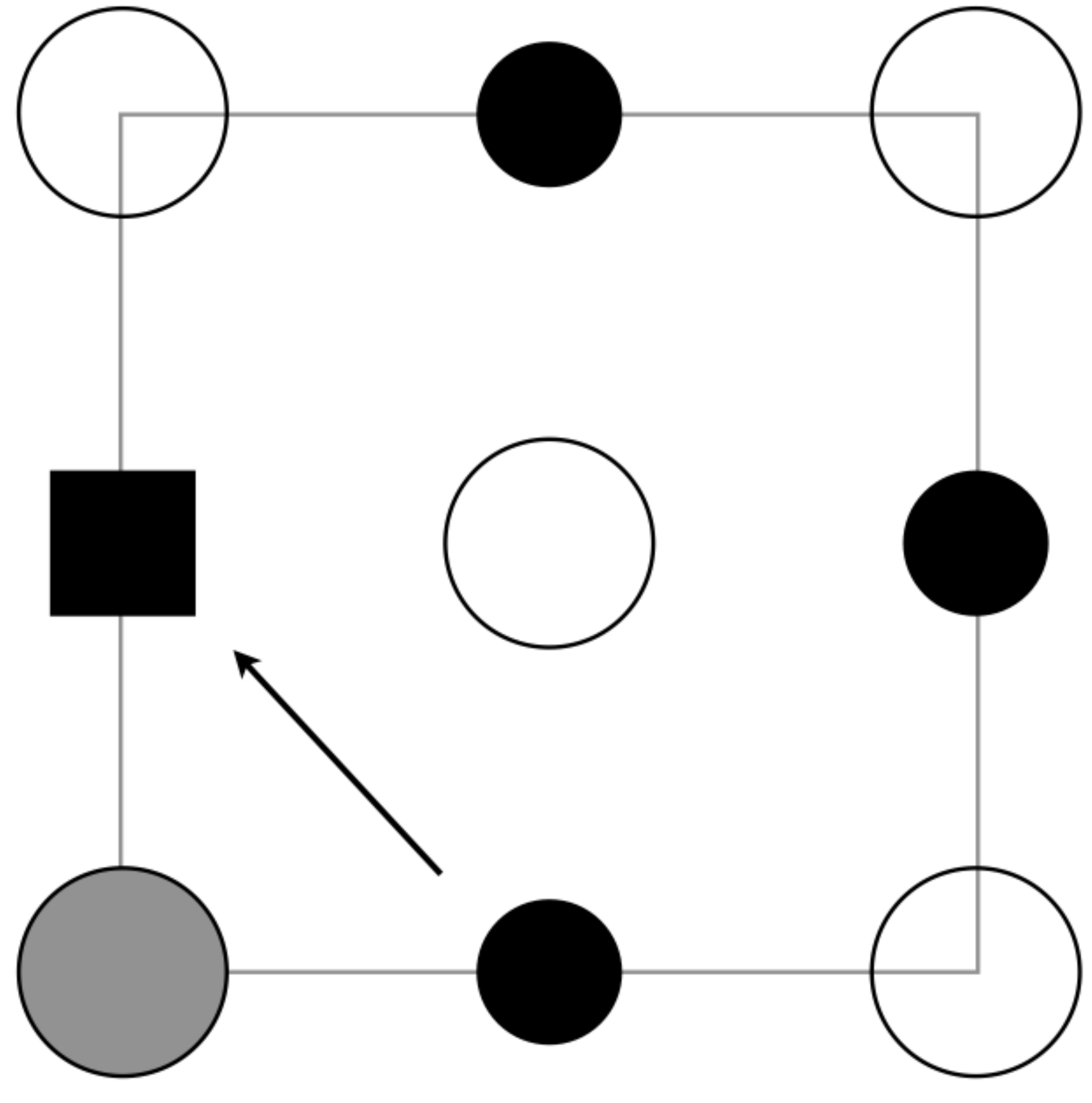}\qquad\includegraphics[width=3cm]{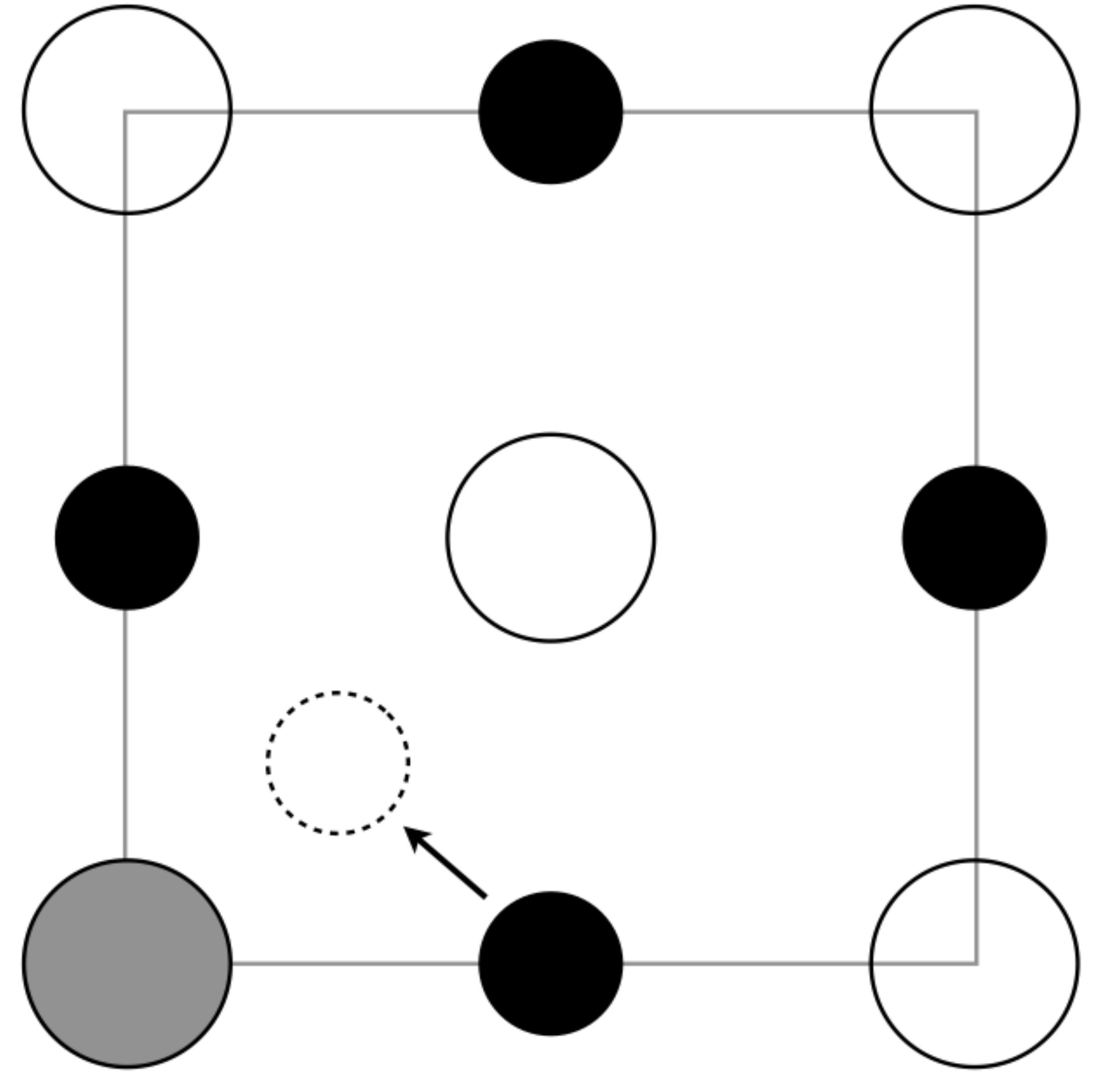}
\caption{\label{fig:barrierillustration} Illustration of the diffusion processes that are investigated in the present study: Vacancy mediated C diffusion (left) and C to interstitial diffusion, i.e. the creation of a type II or Frenkel defect (right). The illustrations show a part of the (100) surface of the B1 structure with metal atoms represented by white and grey circles and C by black circles. Black squares represent C vacancies. Grey circles represent TM impurities. The arrows show the direction of the motion. In case of the C to interstitial process in the right panel the interstitial position is represented by the dashed circle. This position does not belong to the same plane as the other atoms as is clear from Fig.~\ref{fig:defects_ill}.}
\end{figure}

In order to simulate the mobility of various defects, we have calculated the migration energy barriers for diffusion of atoms in TiC. Here we have considered two types of diffusion processes, illustrated in Fig.~\ref{fig:barrierillustration}, that are relevant for C in TiC: The first is vacancy mediated diffusion, in which a C atom moves from a position on the lattice to a nearby vacant position on the C lattice. The second process is related to the formation of the Frenkel type of defect, where a C atom moves from the lattice into a nearby tetrahedrally coordinated interstitial position. The barriers have been calculated using the nudged elastic band method,\cite{neb1,neb2,neb3,neb4} where we have used a minimum of 4 images between the different end points. This has been tested so that the acquired barriers do not change when increasing the number of images.
\par
The defect formation energies and migration energy barriers have been evaluated using the supercell method, where the supercells have been built up by 32 primitive units cells $N$, i.e. a total of 64 atomic positions. This means that if a single TM impurity atom is substituted for Ti, we have an effective concentration of impurities on the metal lattice of $1/32=0.03125$, i.e. about 3~\% of the Ti atoms are substituted. The same concentrations are found when evaluating the C vacancy formation. These cells are large enough to yield reliable trends when it comes to defects formation energies as well as for migration energy barriers. Similar supercell size was also used by Razumovskiy et al.\cite{Volya} in their study of diffusion of intrinsic defects in TiC.
\par
\begin{figure}[t]
\includegraphics[width=7cm]{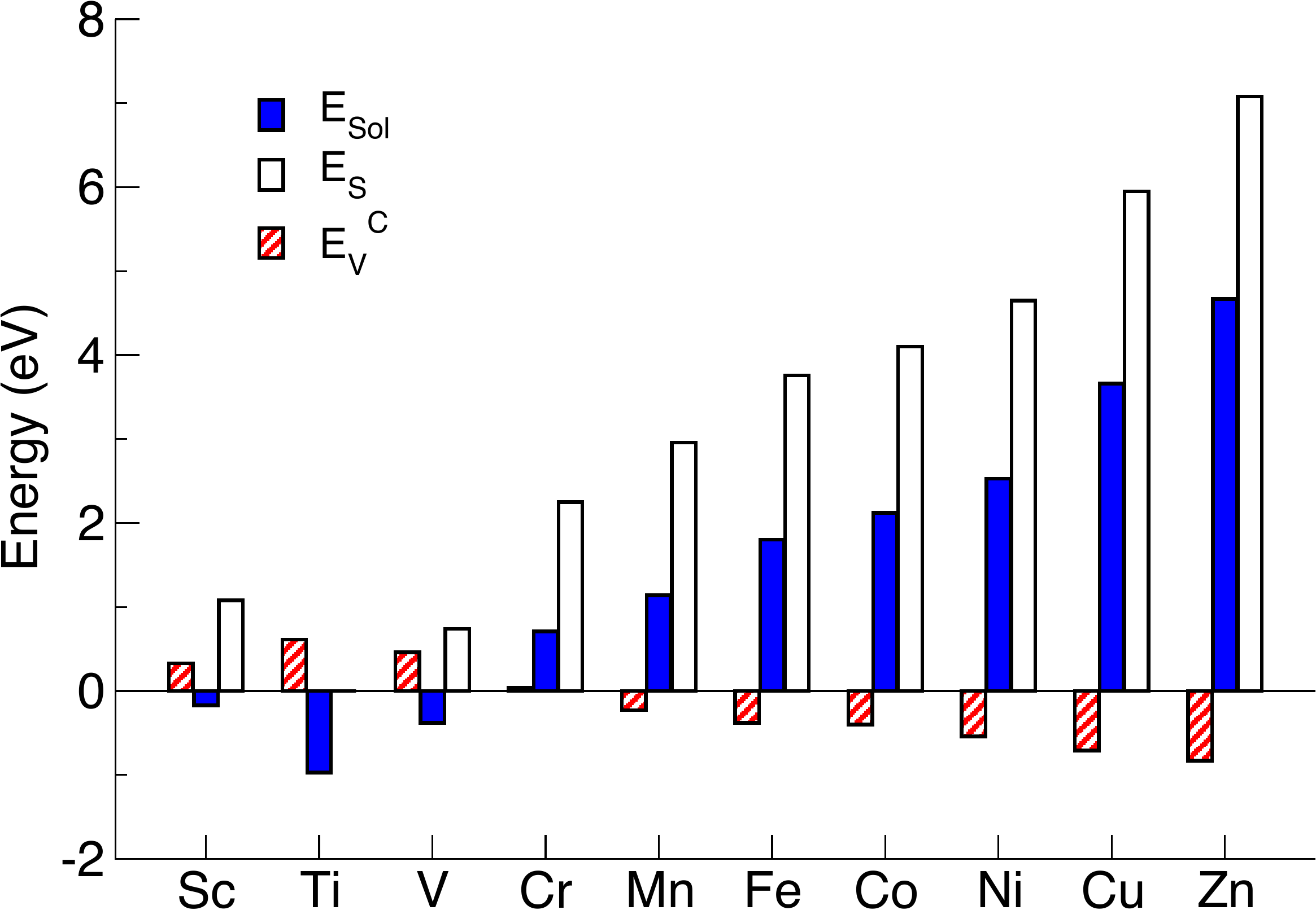}
\caption{\label{fig:defects} (Color online) Evaluated defect formation energies for TiC, where $E_{Sol}$, $E_{s}$, and $E_{V}^{\rm C}$ have been evaluated according to Eqs.~(\ref{eq:dissolved}), (\ref{eq:substitutional}), and (\ref{eq:vacancy_defect}) respectively. The labels on the x-axis marks the TM impurity. Note that Ti represent the case of pure TiC.}
\end{figure}
\section{Results}\label{results}
The ability for TM impurities to be dissolved into TiC is guided by the reaction
\begin{equation}\label{eq:solub}
(N-1){\rm TiC} + {\rm M}_{s} \rightarrow {\rm Ti}_{N-1}{\rm M}_{s}{\rm C}_{N-1},
\end{equation}
where $N-1$ is the amount of TiC and M$_{s}$ is an impurity atom. The expression on the right of Eq.~(\ref{eq:solub}) is equivalent to having a TM atom M$_{s}$ dissolved into TiC by the creation of Ti$_{N-1}$M$_{s}$C$_{N-1}$, i.e. a TM atom is positioned on the Ti lattice followed by the expansion of the lattice by the formation of a C vacancy. We therefore evaluate the energy for dissolving TM impurities into TiC according to
\begin{equation}\label{eq:dissolved}
E_{Sol} = E({\rm Ti}_{N-1}{\rm M}_{s}{\rm C}_{N-1}) - E({\rm Ti}_{N-1}{\rm C}_{N-1}) -E({\rm M}_{s}),
\end{equation}
where $E({\rm Ti}_{N-1}{\rm M}_{s}{\rm C}_{N-1})$ is the total energy of TiC with one substitutional 3d TM defect ${\rm M}_{s}$ and one C vacancy, $E({\rm Ti}_{N-1}{\rm C}_{N-1})$ is the energy of defect free TiC, and $E({\rm M}_{s})$ is the energy of the TM  in its standard reference state.\cite{comment} The result is shown in Fig.~\ref{fig:defects}, where it is found that Sc and V are the only impurities that may be favorably dissolved into TiC. This is because Sc and V are the only metals in addition to Ti that form a TMC in the rock salt structure and our result is in agreement with the results obtained for alloyed solutions of (Ti,M)C in Ref.~\onlinecite{rasander}.
\par
We also define the energy required to substitute a TM atom for Ti by
\begin{equation}\label{eq:substitutional}
E_{s} = E({\rm Ti}_{N-1}{\rm M}_{s}{\rm C}_{N}) - E({\rm Ti}_{N}{\rm C}_{N}) - \left\{E({\rm M}_{s}) - E({\rm Ti})\right\},
\end{equation}
where $E({\rm Ti}_{N-1}{\rm M}_{s}{\rm C}_{N})$ is the total energy of TiC with one substitutional 3d TM defect ${\rm M}_{s}$, $E({\rm Ti}_{N}{\rm C}_{N})$ is the total energy of defect free TiC, and $E({\rm M}_{s}/{\rm Ti})$ is the energy of the metal M$_{s}$/Ti.
In addition, the formation energy of a C vacancy is given by
\begin{equation}\label{eq:vacancy_defect}
E_{v}^{{\rm C}} = E({\rm Ti}_{N-1}{\rm M}_{s}{\rm C}_{N-1}) - E({\rm Ti}_{N-1}{\rm M}_{s}{\rm C}_{N})  + E({\rm C}), 
\end{equation}
where the notation is consistent with Eqs.~(\ref{eq:dissolved}) and (\ref{eq:substitutional}), and $E({\rm C})$ is the energy of one C atom in a graphite unit cell consisting of 4 C atoms. By the use of Eqs.~(\ref{eq:substitutional}) and (\ref{eq:vacancy_defect}) it is possible to rewrite Eq.~(\ref{eq:dissolved}) as 
\begin{equation}\label{eq:dissolved_2}
E_{Sol} = E_{s} + E_{v}^{{\rm C}} + E^{F}({\rm TiC}),
\end{equation}
where $E^{F}({\rm TiC})$ is the formation energy of TiC, i.e. $E^{F}({\rm TiC}) = E({\rm TiC})-E({\rm Ti}) -E({\rm C})$. As is evident from Eq.~(\ref{eq:dissolved_2}), the ability to dissolve TM impurities into TiC depends on the energy required to substitute a Ti atom in TiC for a TM impurity ($E_{s}$), the ability for C vacancy formation in the presence of a TM impurity ($E_{v}^{{\rm C}}$) as well as the formation energy of TiC. The energy required to substitute a TM for Ti and the formation energy of a C vacancy are also shown in Fig.~\ref{fig:defects}. It is clear that the substitution of a TM for Ti is an unfavorable process with a very high formation energy, especially towards the end of the TM series. This behavior is understood by considering a rigid band model where the substitution of Ti for a TM atom with a larger number of valence electrons leads to the occupation of antibonding states which increases the total energy of the system compared to pure TiC, see the discussion in Section~\ref{introduction}. In the case of a Sc impurity, the occupation of binding states is lower compared to the pure TiC which also increases the total energy of the system. The C vacancy formation in the presence of TM impurities is only unfavorable in the case of Sc and V impurities. A transition is made for a Cr impurity after which the C vacancy formation energy becomes negative. This means that in the presence of these latter TM defects, C vacancies will spontaneously be formed. In addition, we note that in the presence of any of the TM defects, the C vacancy formation is lower than in pure TiC, which means that the binding strength between the TM and the surrounding C atoms is weaker than between Ti and C. 
\par
We conclude that the main reason for the difficulty of dissolving TM impurities into TiC stems from the substitutional energy ($E_{s}$) component of Eq.~(\ref{eq:dissolved_2}). The energy required for the substitution of Sc and V for Ti is low enough to be overcome by the formation energy of TiC, resulting in a negative solubility energy $E_{Sol}$. For the other TM defects, the substitutional energy is so large that the strong binding in TiC is not enough to allow for dissolving these metals into the system, even though the C vacancy formation energy becomes negative for the later TM elements. 
\par
In addition, we note that the C vacancy formation energy, given by Eq.~(\ref{eq:vacancy_defect}), corresponds to the formation of a C vacancy by the removal of a C atom in TiC that is then moved out of the carbide and into the form of graphite. This reaction is valid when there is graphite present with which C can be exchanged, e.g. for the case of a carbide phase embedded in a C matrix. Such a system is considered to be rich of C. For Ti rich (C poor) conditions, C vacancies can be formed when Ti is dissolved into the TiC phase, described by the reaction in Eq.~(\ref{eq:solub}) with $M_{s}=$~Ti and evaluated using Eq.~(\ref{eq:dissolved}). This energy is shown in Fig.~\ref{fig:defects}, and it is negative which means that for Ti rich conditions, Ti will be dissolved into TiC with the associated formation of C vacancies. This shows that it is more favorable to have C vacancies in TiC than it is to form a separate Ti phase and can be related to the fact that sub-stocihiometric TiC$_{x}$ with $0.5\leq x\leq1$ is stable against decomposition into Ti and C.\cite{rasander,Hugosson2} We note that the difference between the C vacancy formation energy for Ti rich conditions, evaluated according to Eq.~(\ref{eq:dissolved}), and the C vacancy formation energy for C rich conditions, evaluated by Eq.~(\ref{eq:vacancy_defect}), is identical to the formation energy of TiC, $E^{F}({\rm TiC})$, as can be seen in Eq.~(\ref{eq:dissolved_2}), since in the case of pure TiC the substitutional defect formation energy $E_{s}=0$. C vacancy formation in pure TiC for Ti rich conditions is therefore favorable, while it is unfavorable for C rich conditions. Even so, the evaluated trends in the vacancy formation energy in the presence of TM impurities will vary in the same fashion, irrespective of the external conditions.
\par
\begin{figure}[t]
\includegraphics[width=7cm]{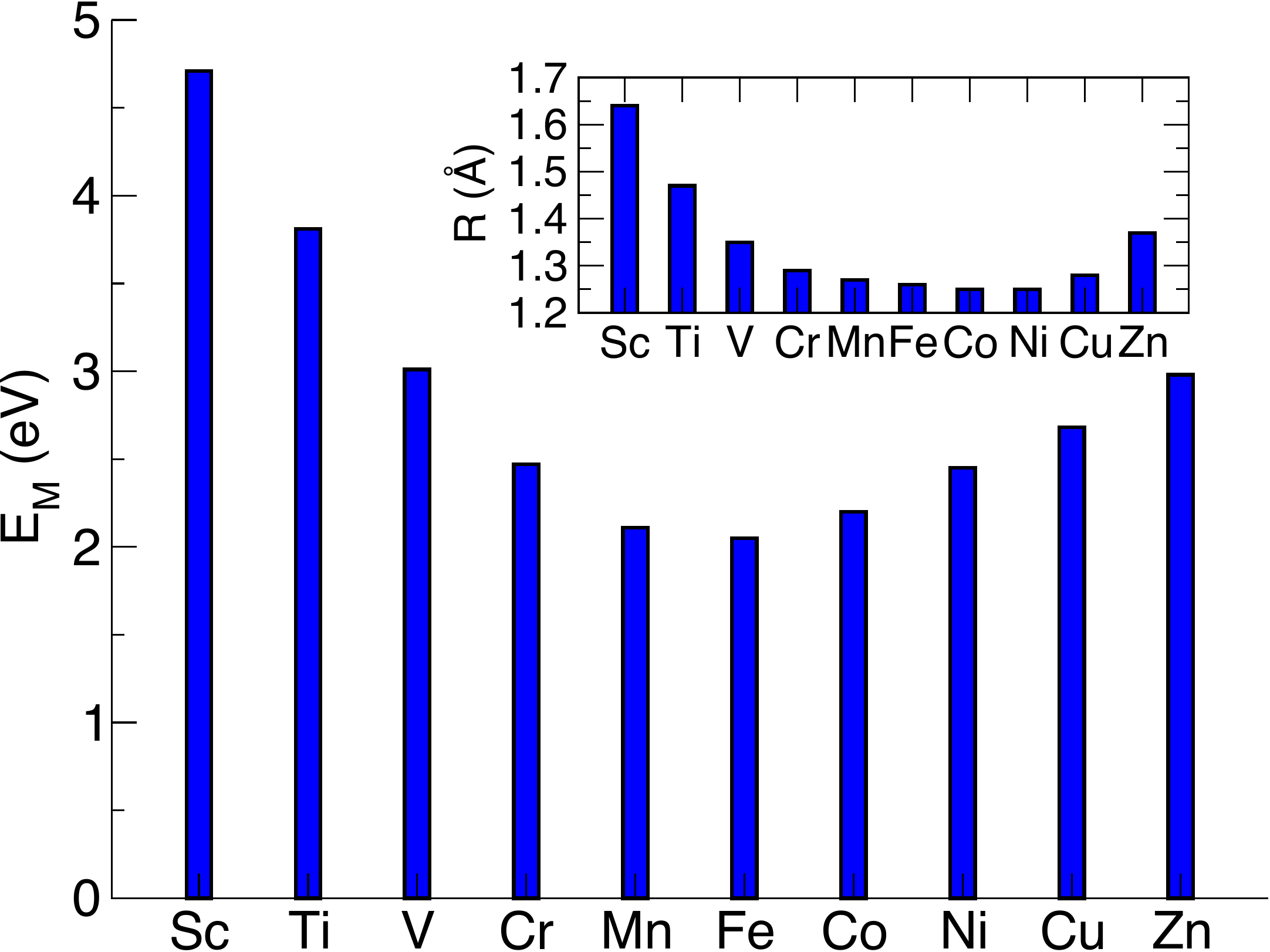}
\caption{\label{fig:vacancybarrier} (Color online) The calculated C vacancy mediated migration energy barriers $E_{M}$ in the presence of TM impurities. Note that Ti represent the case of pure TiC. The inset shows the metallic radius $R$ of the TM atoms. The metallic radii given here are taken from Ref.~\onlinecite{Wells}.}
\end{figure}
In Fig.~\ref{fig:vacancybarrier} we present the evaluated vacancy mediated migration energy barriers of C in the presence of TM impurities. The value obtained for this process in pure TiC (second staple from the left in Fig.~\ref{fig:vacancybarrier}) is in agreement with the value obtained by Tsetseris et al.\cite{Tsetseris} of 3.46~eV. The difference between our value and the value of Tsetseris et al. is attributed to the use of different sized supercells. Our study is done using a smaller supercell, which will yield a larger value for the migration energy barrier compared to a larger supercell, since atoms are allowed to relax further in a larger supercell and thereby lowering the energies along the migration path. The migration energy barrier also correspond rather well with experimental observations (3.30~eV,\cite{vanLoo} 3.60~eV,\cite{Kohlstedt} 4.15~eV\cite{Ettmayer}). We note, however, that lower barriers have been obtained experimentally in the order of 2.44-2.70~eV\cite{Kohlstedt} but these have been related to a different diffusion process by Tsetseris et al.\cite{Tsetseris} where the C atoms migrate between interstitial sites and not between lattice sites on the C lattice as has been evaluated here. 
\par
According to the data shown in Fig.~\ref{fig:vacancybarrier}, the vacancy mediated process has a higher barrier if it takes place in the presence of Sc impurities compared to pure TiC. All other TM impurities yield a smaller barrier. We also note a clear trend along the TM series: The highest values are found in the beginning and the end of the series, with a clear minimum in the migration energy barrier in the middle. This trend does not reflect the same behavior as the stability of the TM impurities, neither does it reflect the stability of the C atoms in the vicinity of the TM impurities, see Fig.~\ref{fig:defects}. Rather, we find that the behavior of the migration energy barriers reflects the behavior of the size of the TM involved. In the inset of Fig.~\ref{fig:vacancybarrier}, we show the metallic radius of the TM atoms. As can be seen in Fig.~\ref{fig:vacancybarrier}, the metallic radius along the 3d TM series has its largest value for Sc, with a minimum in the middle of the series, and a low rise at the end. The finding that the height of the migration energy barriers is related to the atomic sizes is reasonable since a large impurity atom will inhibit the motion more than a smaller impurity, which is exactly what is shown in Fig.~\ref{fig:vacancybarrier}. In the past, the size of the TM impurities have also been found to correlate with other materials properties, e.g. the tendency for surface segregation of TM impurities of the TiC(100) surface.\cite{Hugosson1}
\begin{figure}[t]
\includegraphics[width=7cm]{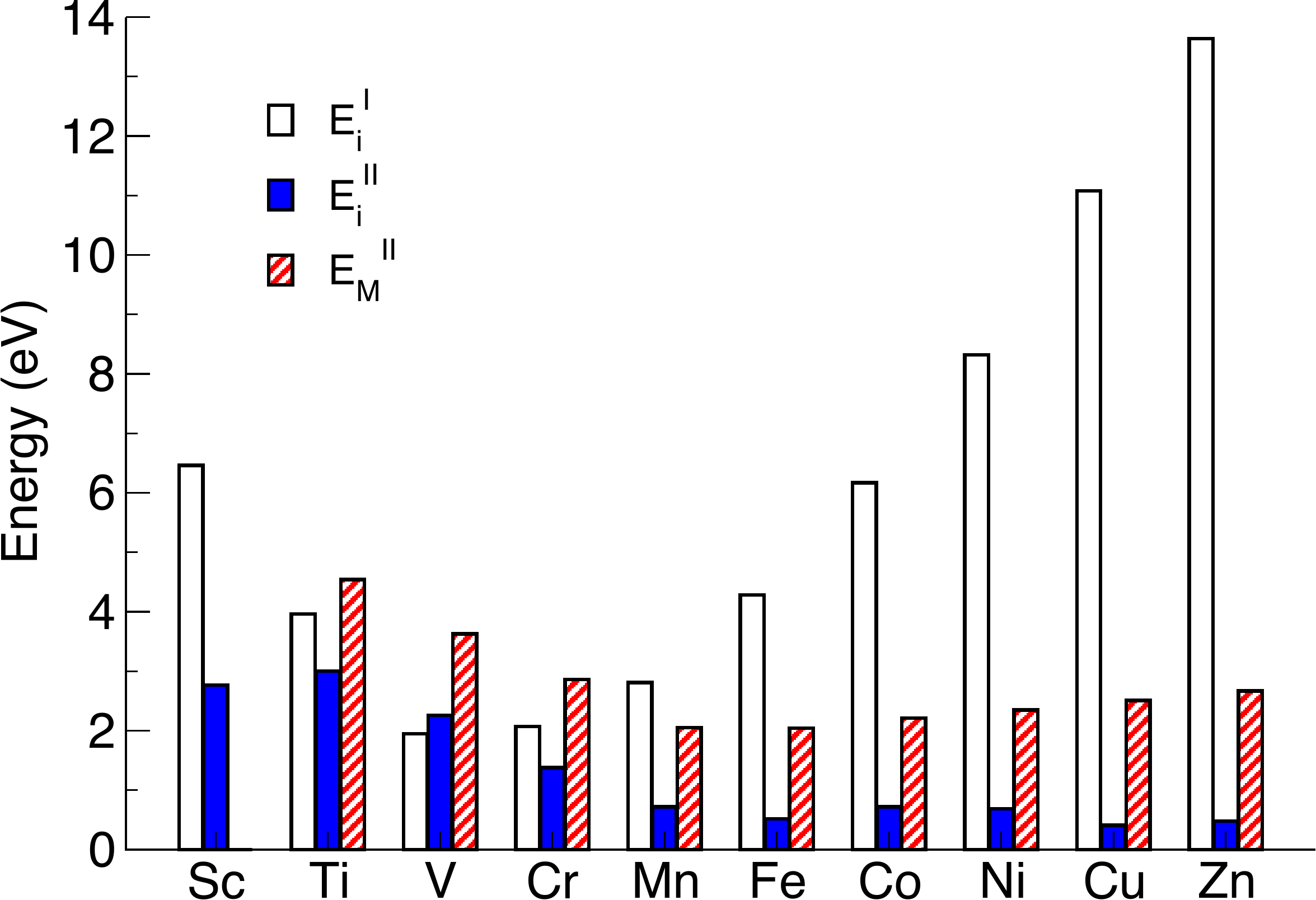}
\caption{\label{fig:interstitial} (Color online) The calculated formation energies of type I and II C interstitial defects according to Eqs.~(\ref{eq:typeI}) and (\ref{eq:typeII}). The migration energy required to from the type II defect, $E_{M}^{II}$ is also shown. Note that Ti represent the case of pure TiC and, in addition, the migration barrier in the presence of Sc impurities is not evaluated.}
\end{figure}
\par
We now turn the attention to C interstitial defects and the results are shown in Fig.~\ref{fig:interstitial}. The interstitial defect of type I in the presence of TM impurities is evaluated as
\begin{equation}\label{eq:typeI}
E_{i}^{I} = E({\rm Ti}_{N-1}{\rm M}_{s}{\rm C}_{N+1}) - ({\rm Ti}_{N-1}{\rm M}_{s}{\rm C}_{N}) - E({\rm C}),
\end{equation}
where $E({\rm Ti}_{N-1}{\rm M}_{s}{\rm C}_{N+1})$ is the energy of a system with an additional C atom in the interstitial position $I$ in Fig.~\ref{fig:defects_ill}, while the interstitial defect of type II is evaluated as
\begin{equation}\label{eq:typeII}
E_{i}^{II} = E({\rm Ti}_{N-1}{\rm M}_{s}{\rm C}_{N}^{\ast}) - E({\rm Ti}_{N-1}{\rm M}_{s}{\rm C}_{N}),
\end{equation}
where $E({\rm Ti}_{N-1}{\rm M}_{s}{\rm C}_{N}^{\ast})$ is the energy of a system with a C atom at the interstitial position $I$ and a vacancy at position $V_{I}$ in Fig.~\ref{fig:defects_ill}. We find that the formation energy of a type I C interstitial defect (shown in Fig.~\ref{fig:interstitial}) has a minimum in the presence of V and Cr defects and that this energy becomes very large towards the end of the 3d TM series. This behavior is also related to the size of the TM elements, but also strongly related to the TM elements ability to form TMC which is strongly reduced towards the end of the TM series. The type I defect has an additional C atom in the tetrahedrally coordinated interstitial site next to a TM atom and it is not surprising that the formation energy is high and that it shows a clear atomic size contribution since the density of atoms is large. The type II defect, or Frenkel defect, on the other hand has a much lower formation energy. In Fig.~\ref{fig:interstitial}, we see that the type II defect is easier to form close to TM impurities compared to the pure TiC and towards the end of the TM series the formation energy becomes rather small. 
\par
In Fig.~\ref{fig:interstitial}, we also show the migration energy barriers for moving a C atom on the lattice, $V_{I}$ in Fig.~\ref{fig:defects_ill}, to the interstitial site $I$ in Fig.~\ref{fig:defects_ill}. We find that this barrier has a similar behavior in the presence of TM impurities as the C vacancy mediated migration barrier shown in Fig.~\ref{fig:vacancybarrier}, which is again related to the sizes of the TM impurities compared to Ti. 
\par
\section{Summary and conclusions}\label{conclusions}

In summary, we have performed density functional calculations of the formation energies of TM impurities, C vacancies, and C interstitial defects in TiC. We find that the substitution of TM elements for Ti is an unfavorable process and, in fact, the only elements that can be favorably dissolved into TiC at equilibrium are Sc and V. However, it has been found that during non-equilibrium conditions large quantities of TM, e.g. Fe, Ni, Pt and Al, can be dissolved into TiC to form ternary TMC systems.\cite{wilhelmsson,rasander,Jansson,Jansson2,refB4,refG1,refG2}  Furthermore, we find that the presence of TM impurities greatly affects the stability and mobility of C atoms in the vicinity of the TM.  Especially, the C vacancy formation energy is always lower in the presence of the impurities compared to pure TiC, and therefore the impurities will attract C vacancies. We also find that the migration energy barriers of C are always lower in the presence of TM impurities compared to pure TiC, except for the case of Sc, due to the atomic size effect of the TM atoms compared to Ti: If the size of the TM is smaller than Ti, the migration energy barrier will be lower. This means that the mobility of C is higher in systems with large quantities of TM impurities, and especially so for TM impurities from the middle of the 3d TM series. We therefore propose that the C release  effect shown to exist for alloyed solutions of (Ti,M)C\cite{wilhelmsson,rasander,Jansson,Jansson2} will be more favorable if the alloying element is taken from the middle of the 3d TM series.

\section{Acknowledgements}
We acknowledge financial support from the Swedish Research Council, Swedish Foundation for Strategic Research and the Carl Trygger Foundation. O. E. is grateful to the European Research Council (project 247062 - ASD) and the KAW foundation for support.  The calculations were performed on resources provided by the Swedish National Infrastructure for Computing (SNIC) at High Performance Computing Center North (HPC2N) and at Uppsala Multidisciplinary Center for Advanced Computational Science (UPPMAX).


\begin{thebibliography}{999}
\bibitem{comment} The reference states that have been used are presented in Ref.~\onlinecite{rasander}.
\end{thebibliography}
\end{document}